\documentstyle[12pt,epsf]{article}

\begin{document}

\font\sansf=cmss12

\begin{titlepage}

\rightline{DAMTP-R95/14}
\vspace{0.2in}
\LARGE
\center{Numerical Solutions to  Dilaton Gravity
Models and the Semi-Classical Singularity}
\Large
\vspace{0.4in}
\center{}
\center{J. D. Hayward\footnote{E-mail address:
J.D.Hayward@damtp.cam.ac.uk}}
\vspace{0.3in}
\large
\center{\em Department of Applied Mathematics and Theoretical Physics,
\\ University of Cambridge,
\\  Silver Street, Cambridge CB3 9EW, U.K.}

\vspace{0.4in}
\small\center{ }
\vspace{0.5in}
 A general homogeneous two dimensional dilaton
gravity
model considered recently by Lemos and S\` a
\cite{LEMSA}, is given quantum matter Polyakov corrections and
is solved numerically for several static, equilibrium scenarii.
Classically the dilaton field ranges the whole real line, whereas
in the semi-classical theory, with the usual definition, it is
always below a certain critical value\cite{BGHS,SHE} at which a
singularity appears. We give solutions for both
sub- and super-critical dilaton field.
The pasting together of the spacetime on both sides of a
 singularity in semi-classical planar general relativity is
discussed.

\vspace{0.5truecm}

\setcounter{footnote}{0}
\end{titlepage}

\baselineskip=0.2in
\section{Introduction}
\setcounter{equation}{0}

The machinations of black holes have been studied extensively in
recent years
using two dimensional models \cite{CGHS,RST}.

In\cite{BGHS,SHE},the original {\sl CGHS} model was solved numerically
for
explicitly static equilibrium scenarii. The {\sl CGHS} lagrangian
is a
dilaton gravity model which is made up of a classical part, which
comes directly from low energy string theory in two dimensions,
and a quantum correction described by Polyakov in\cite{POL} which
takes into account the one-loop effects of the matter fields. The
number of these fields can be proliferated so that the effect of
other quantum corrections is small compared to that of the matter.
In another paper, by Lemos and S\` a\cite{LEMSA}, the classical
lagrangian considered is more general than that of the {\sl CGHS}
 model.
A variable multiplicative parameter is included in the kinetic
dilaton term. By choosing certain values for this parameter, a
set of classical models which includes low energy string theory
in two dimensions, Jackiw-Teitelboim theory, and planar general
relativity is obtained.

In this paper, the idea is to combine and extend the work of
\cite{LEMSA} and \cite{BGHS,SHE}.
The more general classical core lagrangian of \cite{LEMSA}
will be combined with the Polyakov quantum matter correction, and
a set of static numerical solutions to various models with and
without the correction are displayed.
The static black holes in equilibrium with Hawking radiation can be
studied numerically.
One motivation for studying such solutions is to understand the
`quantum' singularity which
was discovered shortly after the appearance of the original
{\sl CGHS}
 paper\cite{RSTA,BAN}. Birnir {\it et al}  noted that this
singularity occurred at the finite dilaton value, and that the
metric was actually finite there, unlike the classical
case. We would like to investigate this further here.
 The static
solutions might be a candidate final state of black
hole evaporation, {\it i.e.} as massive remnants.
This was rejected
in\cite{BGHS} since the ADM mass for these solutions is
divergent because there is non-zero radiation density out to
infinity. Here we shall find the expression for the
ADM mass in that case and show that it is indeed infinite.
For equilibrium in two dimensions, this divergence is actually
necessary, but we shall see that the solutions are nevertheless
interesting.

 In the following section,
a general two dimensional homogeneous dilaton gravity model is
introduced, whose field equations for static solutions are
written down.  The initial
conditions for regular-horizon spacetimes are then given.

In section three,
 a general introduction to the results
is given.
Static regular horizon solutions in which the dilaton is
initially and
remains sub- or super-critical are then found for a range of
classical cores with and without quantum corrections.
The solutions being static, and numerical, it
is difficult to be precise about global structure away
from the singularities at infinity, though one can
give some details about the singularity itself and the
horizon.
 We restrict a
fuller description and interpretation to
the case whose classical core is that of planar general
relativity.

In section four are the conclusion and discussion.

\section{General Homogeneous Two-dimensional Model}
\setcounter{equation}{0}
\label{sect:models}
\subsection{Introduction}
 A general homogeneous lagrangian with
semi-classical minimal
scalars is of the form
\begin{equation}
I = \frac 1 {2\pi}\int d^2 x \sqrt{\pm g}
\Big[ R\tilde\chi(\Phi)+4\omega e^{-2\phi}{\nabla \phi}^2+V(\Phi)
-\frac {\kappa} 4 {\nabla Z}^2-\frac 1 2 \sum_{i=1}^N(\nabla f_i)^2 \Big]
-\frac 1 {\pi} \int d \Sigma \sqrt{\pm h} K\tilde\chi, \label{eq:gen}
\end{equation}
where $\tilde\chi=\frac {\kappa} 2 Z + \Phi,$ and $\Phi$ is a function
of the dilaton
field $\phi$ and any other fields that are required.

The terms involving Z in the volume term of this action have replaced
the
usual Polyakov term\cite{POL}
$\frac {\kappa} 4 R(x) \int d^2 x' \sqrt{-g(x')} G(x,x')R(x') $
which comes from the matter contribution to the
associated path integral.  $G(x,x')$ is the scalar Green's function.
The trace anomaly of the Z scalar field is
that of the N minimal scalars.

 One could choose the
function of the dilaton field so as to make the theory have
vanishing
central charge\cite{ABC}, but for simplicity models for which
 $V(\Phi)=4\lambda^2,$
 will be
restricted to here. The function $\tilde\chi$ takes into account
both
 the classical
coupling of the dilaton to gravity, and any one-loop terms which
come
from quantising
additional fields. The classical part will be taken to be
$e^{-2\phi},$
and the one-loop corrections to be those of the
{\sl CGHS} model, so that
$\tilde\chi=e^{-2\phi}-\frac {\kappa} 2 (\epsilon\phi-Z)$,
where $\epsilon=0$
\footnote{The {\sl RST} \cite{RST}
model has $\epsilon=1,\omega=1$.}.
The {\sl CGHS} model is regained when $\omega=1.$

If one tries to work in the two-dimensional analogy of
the `Eddington-Finkelstein' gauge\cite{JH2},  the action and field
equations are still complicated although the work on entropy
which depends on the position of the horizon might be
simplified.
However for static solutions one should use the conformal gauge
where the line element is  \begin{equation}
dl^2=-e^{2\rho}dx^{+}dx^{-}.  \label{eq:conf1}
\end{equation} The field equations then imply that $Z=2\rho$,
 up to a solution of the wave equation.

The action (\ref{eq:gen}) now becomes
\begin{equation}
I=-\frac 1 {\pi} \int d^2 x \sqrt{g} \Big[e^{-2\phi}
(2\partial_- \partial_+ \rho  -4\omega\partial_- \phi\partial_+ \phi+
\lambda^2e^{2\rho} )+
\frac 1 2 \partial_+  f_{i}\partial_-  f_{i}
-\kappa(\partial_- \rho\partial_+ \rho +
\epsilon\phi\partial_- \partial_+ \rho) \Big] \label{eq:action}
\end{equation}
The surface term is omitted from now on.

\subsection{Field Equations}
\label{sect:fe}
The following applies to static solutions which are functions of
the static variable $s=-x^+x^-$
only.
 Terms in
$f_i$ have been set to zero.
 Denoting
$'=\frac d {ds}$, the field equations become

\begin{equation}
Q(\phi) [\phi '' +\frac 1 s  \phi '] =
2{\phi '}^2(P(\epsilon,\phi)-\frac 1 2 \omega\kappa e^{2\phi})-
\frac {\lambda^2} {2s}e^{2\rho}(P(\epsilon,\phi)-\frac 1 2 \kappa e^{2\phi})
\label{eq:field1}
\end{equation}
\begin{equation}
Q(\phi) [\rho '' +\frac 1 s  \rho '] =
2\omega{\phi '}^2(2-P(\epsilon,\phi))-\frac {\lambda^2}
{2s}e^{2\rho}(2\omega-P(\epsilon,\phi))
\label{eq:field2}
\end{equation}
where $P(\epsilon,\phi)=1+\frac {\epsilon\kappa} 4 e^{2\phi}$ and
$Q(\phi)=P(\epsilon,\phi)^2-\omega\kappa
e^{2\phi}$.

The constraint equations are
\begin{equation}
 P(\epsilon,\phi)(\phi '' -2\rho '\phi ')
+2(\omega-1){\phi '}^2 +   \frac 1 2 \kappa e^{2\phi} ({\rho '}^2   -\rho ''+
\frac t
{s^2})=0  \label{eq:con}
 \end{equation}
where t is given by the boundary conditions required.
 These
equations which will reduce in the case of $\omega=1$ to either the CGHS
 ($\epsilon=0$) or
RST ($\epsilon=1$) model.

\subsection{Initial Conditions}

One can solve the dynamical equations numerically for $\rho,$ $\phi$ and their
derivatives, by rewriting them as a coupled set of four first order
differential
equations.  The boundary conditions chosen are such that the origin
 in $s$ is regular.
This requires that
\begin{equation}
s\rho''(0)=s\phi''(0)=0, \label{eq:cdn1}
\end{equation}
\begin{equation}
 t=0, \label{eq:cdn2}
\end{equation}
A shift in $\rho$ allows one to remove $\lambda$ from the equations.
 $\phi$ can also be
redefined so that the equations are independent of $\kappa$.  One then
finds that the
derivatives at the origin should be
 \begin{equation}
\rho '(0)=\frac {-\frac 1 2
e^{2\rho_{0}}[P(\phi_0)-2\omega]} {Q(\phi_0)}
\label{eq:dero}
\end{equation}
\begin{equation}
\phi '(0)=\frac {-\frac 1 2
e^{2\rho_{0}}[P(\phi_0)-\frac 1 2  e^{2\phi_{0}}]} {Q(\phi_0)}
\label{eq:derphi}
\end{equation}
These equations reduce to those of \cite{BGHS} in the case $\omega=1,
\epsilon=0.$

Varying the initial value of $\rho$ simply scales the equations.
The initial value of $\phi$ at the origin is related to the `size'
of the black hole. That is, moving towards the critical value initially,
reduces the coordinate distance to the singularity from the horizon.

\subsection{Choice of $\omega$-Parameter}
\label{sect:which}
In order to choose the smallest set of values of
$\omega$ each of
 which produce
different behaviour,
one can consider the expression
(\ref{eq:rq}) for the
curvature, $R=-8e^{-2\rho}\frac {d} {ds}+s\frac {d^2} {ds^2}\rho$.
 the field equations
(\ref{eq:field1},\ref{eq:field2})
and the values of $\omega$
and $\phi_0$ which change the sign of the initial values of
$\frac {d\rho} {ds}$ and
$\frac {d\phi } {ds}$. The critical value of $\phi$
\footnote{$\omega$ is multiplied by $\kappa$ in the logarithm
 if $\kappa$ has not been scaled out.}   is given by
 \begin {equation}
\phi_{cr}=-\frac 1 2 \log\omega.
\end{equation}
 Singularities will therefore
occur at increasingly weak coupling as $\omega$ is increased.
The CGHS model(see section \ref{sect:models}.1) corrections are used
 in the following,
{\it i.e.} $\epsilon=0$ and so $P(\phi,\epsilon)=1$.
 The initial value of
$\frac {d\phi } {ds}$
is then zero at $\phi_0=\frac 1 2 \log 2$, unless $\omega=\frac 1 2 $.

One can divide the cases first into those for which $\omega>\frac 1 2 $,
$\omega=\frac 1 2 $, and
$\omega<\frac 1 2 $. Then for the former case one has
$\phi_0<\phi_{cr}$,$\phi_{cr}<\phi_0<\frac 1 2 \log 2$ and
$\phi_0>\frac 1 2 \log 2$. When $\omega<\frac 1 2 $,
we have $\phi _0<\frac 1 2 \log 2 $,  $\frac 1 2 \log 2<\phi_0<\phi_{cr}$
and  $\phi_0>\phi_{cr}$. For $\omega=\frac 1 2$, one simply has
sub- and super-critical initial values, $\phi_0<\phi_{cr}$ and
 $\phi_0>\phi_{cr}$.

 One
can also compare with the classical counterparts for which $\kappa=0$
in the field
equations. The critical values of $\phi$ do not exist in the
classical case.

By inspection of the field equations
(\ref{eq:field1},\ref{eq:field2}), the regions from
which one would
like to consider a value of $\omega$ are
\begin{equation}
{\omega<0; \omega=0; 0<\omega<\frac 1 2 ; \omega=\frac 1 2 ;
\frac 1 2 <\omega<1; \omega=1; \omega>1}
\end{equation}
Lemos and Sa\cite{LEMSA} also show that the
global structure differs
 for the cases
$1<\omega<2$,$\omega=2$,and $\omega>2$. The
numerical analysis does
not distinguish qualitatively between these cases.

We leave until later a fuller discussion of the
case $\omega=\frac 1 2$.

\newpage
\section{Solutions}

 Since the corrected solutions are static by ansatz,
and there is in general non-zero radiation density outside
the black hole due to the Polyakov term associated with the
minimal matter fields, they
represent equilibrium scenarii.  The ADM mass may
be calculated as follows\cite{W}.

Let
 $g_{ab}=\eta_{ab}+h_{ab}$ , and $\phi=\Phi_{L}+\varphi$,
be perturbations from flat space
 $\eta_{ab}$,
and from linear dilaton $\Phi_{L}$, where $h_{ab}$ and $\varphi$
vanish at infinity.
The total mass measured by an observer at right infinity
is given by
\begin{equation}
 M = \int t_{0\mu}\xi^{\mu} dx \label{eq:mass}
\end{equation}
where $t_{0j}$ comes from the linearised energy-momentum tensor
for the classical theory,
$\xi^j$ is a timelike Killing vector, and $x$ is a suitable radial
coordinate.
For this calculation, one needs the generalised asymptotic
expansions of $\rho$ and $\phi$. In the case $\omega=1$, we have
these expressions\cite{SHE}. Below, we shall note the result for
this case, which is representative of the $\omega>0$ cases.
It would seem that $M\to\pm\infty$ except when $\omega=0$. This
is due to the thermodynamics
peculiar to two dimensions.

There is
by construction, an horizon on $s=0$ in all the cases.
For example, in the
case $\omega=1$, when one reproduces the classical black hole of
Witten\cite{W},
there is
a curvature singularity at finite negative $s$, behind the horizon
at the origin.
 One can see how the
distance from the origin to the singularity decreases as one goes
toward the critical
value of dilaton. This corresponds to a smaller black hole, which
would appear later in a
sequence of static black holes that one might use to represent black
 hole evolution.
However, the sequence can never be complete
because of the
divergences as one approaches the critical value.

\subsection{Classical Solutions}

There will be given a set of plots of the numerical regular-horizon
solutions of the
model with the various values of the parameter $\omega$ considered
in \cite{LEMSA} for the
classical case which has $\kappa=0$. For $\kappa=0$, one can show that there
 exists a timelike
killing vector,  so the most general solution is static
\cite{GIBPERB}.

In the classical case, the initial value of the gradient of $\phi$
is given by
\begin{equation}
\phi '(0)=-\frac 1 2   e^{2\rho_{0}}        \label{eq:derphii}
\end{equation}
and $\rho_{0}=0$ is chosen in each case. This initial value is
independent of $\omega$.
For the initial gradient of $\rho$
\begin{equation}
\rho '(0)=-\frac 1 2  (2\omega-1) e^{2\rho_{0}} \label{eq:derrhoi}
\end{equation}
which clearly depends on $\omega$ and goes through zero at
$\omega=\frac 1 2 $.
Let the operator
\begin {equation}
D=\frac {d} {ds}+s\frac {d^2} {ds^2}. \label{eq:D}
\end{equation}
The classical equations are
\begin{equation}
D \phi = 2s{\phi '}^2-\frac 1 2  e^{2\rho},
\label{eq:field1c}
\end{equation}
\begin{equation}
D \rho = 2s\omega{\phi '}^2-\frac 1 2  e^{2\rho}(2\omega-1).
\label{eq:field2c}
\end{equation}

Notice that $\phi$ only appears in the equations as a derivative of
s. This means that
it will not matter as far as qualitative changes are concerned what
 the initial value of
$\phi$ is: there are no {\it critical} values of $\phi_0$.
The initial conditions given above are applied, which ensure
that the solution is
regular at the horizon, $s=0$.
The initial value of $\rho$ simply scales, and is taken in every
case to be zero.
The coordinates $x^+,x^-$ are
analogous to the Kruskal
coordinates in the Schwarzschild solution, and cover the extended
manifold.  The two
coordinate invariant functions $\phi$ and the curvature scalar R,
are plotted, along with
 the metrical factor
$\rho$ for several values of $\omega$.

In order to obtain solutions which are regular at the origin, one
has to integrate
from the origin in both directions using particular initial
conditions on $\rho$ and
$\phi$ and their derivatives.
 The key points are
to note singularities, or lack of singularities in the curvature and
 whether they
occur at weak or stong coupling in $\phi$, and also to note
divergences in $\rho$
and/or $\phi$, whilst the curvature is finite.
Further physical conclusions are difficult to make since these
are numerical solutions which are static, and thus effectively
one-dimensional. The following gives the legend for the numerical
solution plots:

\vbox{\[
\begin{array}{c}
\epsfxsize=0.35\displaywidth
\epsfbox{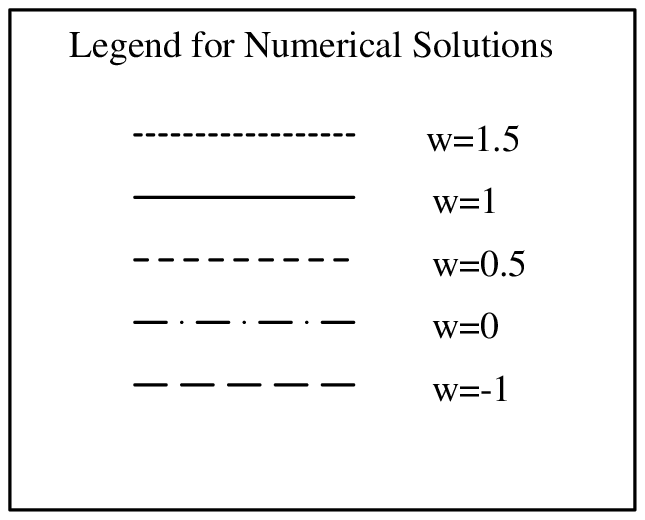}\\
\end{array}\]}

  Our classical numerical
results  are in
agreement with the analytical solutions of \cite{LEMSA} for their
$A>|B|$, with $\lambda^2>0$. It is necessary to reproduce these so
that one can compare with the new semi-classical solutions given
later. In the following, brief physical comments are made upon each
of the solutions, which should be read in conjunction with FIGS.
1-3.

\vspace{0.3in}
\centerline{\bf A. \ $\omega<0$}
\smallskip
We take for example $\omega=-1$. Towards positive $s$,
$\phi$ diverges to
minus infinity, whilst $\rho\to\infty$, at finite coordinate
distance. The curvature is
approximately constant and negative. Thus we have a timelike right
infinity, at weak
coupling. To the left we have $\phi\to\infty$, whilst
$\rho\to-\infty$. The curvature
goes to $-\infty$ so we have a timelike singularity. The
extended manifold is given in \cite{LEMSA}, as for all the following
 classical cases.

\smallskip

\vbox{
\[
\begin{array}{c}
\epsfxsize=0.6\displaywidth
\epsfbox{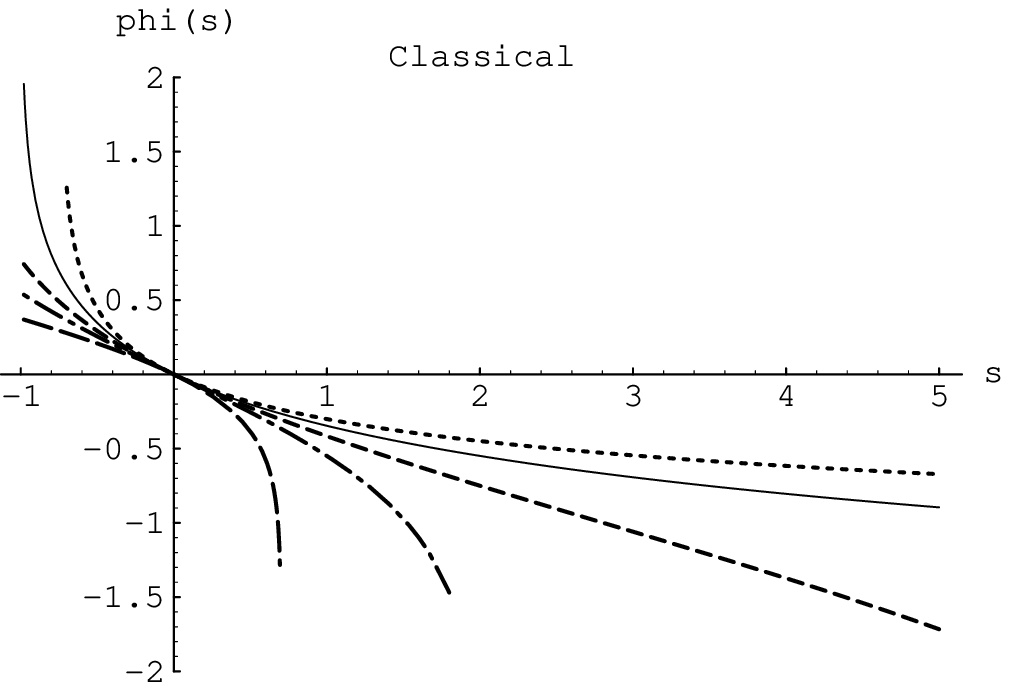}\\
\end{array}\]
FIG. 1. The dilaton field for the five cases of $\omega$-parameter
at the classical level.}

\vspace{0.3in}
\centerline{\bf B. \ $\omega=0$ }

This is the Jackiw-Teitelboim theory  considered in \cite{JT}.
It represents constant curvature anti-de Sitter space with strong
coupling to the
left, and weak coupling to the right.
The second field equation above (\ref{eq:field2c}) is precisely the
 statement that the
curvature scalar will be $R=-4$, and this is borne out in
FIG. 3.

\vspace{0.3in}
\centerline{\bf C. \ $\omega=\frac 1 2 $ }
\smallskip
This is planar general relativity.
The gradient of $\rho$ is initially zero. In the above
cases it is positive, and  becomes increasingly negative as $\omega$
increases.
 There is a spacelike singularity at strong
coupling($\phi\to\infty$) to the left as in the previous case and
for all $\omega>0$. To the right, $\phi$
diverges to minus infinity, and the curvature tends to a constant
negative value.
There is thus a black hole with a timelike right infinity.

\vbox{\[
\begin{array}{c}
\epsfxsize=0.6\displaywidth
\epsfbox{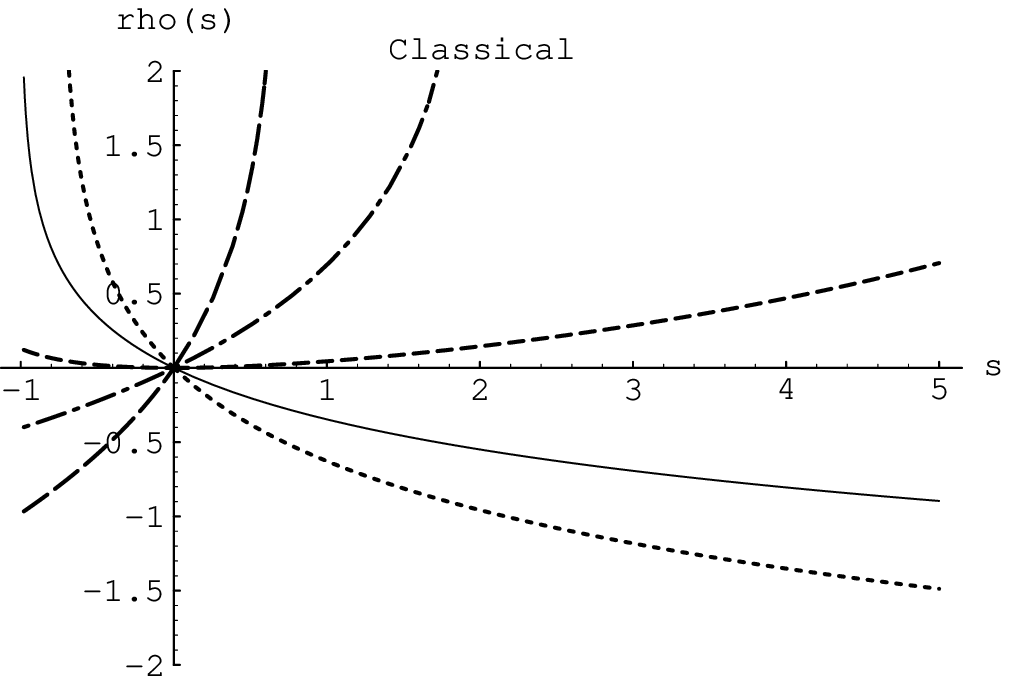}\\
\end{array}\]
FIG. 2. The conformal factor for the five cases of
$\omega$-parameter at the classical level.}

\vspace{0.3in}
\centerline{\bf D. \ $\omega=1$ }
\smallskip
This is the classical black hole of Witten\cite{W}, which was
found in low energy string theory. There is a
spacelike singularity at
strong coupling($\phi\to\infty$), and the curvature tends to zero
at right infinity, which is null.

\vspace{0.3in}
\centerline{\bf E. \ $\omega>1$}
\smallskip
We consider $\omega=\frac 3 2$. This has the
spacelike singularity
at left infinity at strong coupling($\phi\to\infty$), but at right
infinity,  $\phi$,
$\rho$  and the curvature all go to minus infinity logarithmically.
 The rate at which the
curvature does so increases as $\omega$ increases. This means that
these spacetimes
have singularities to the right and left, spacelike and timelike
respectively.  More details can be found in \cite{LEMSA}.

\vbox{
\[
\begin{array}{c}
\epsfxsize=0.6\displaywidth
\epsfbox{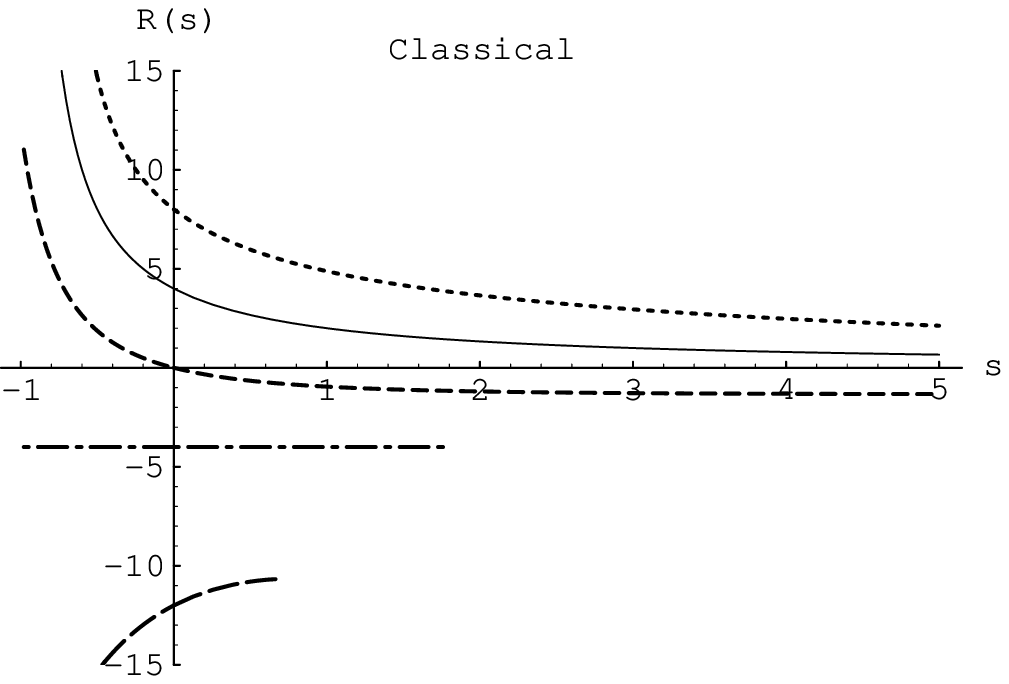}\\
\end{array}\]
FIG. 3. The scalar curvature at the classical level.}

\subsection{Quantum Corrections}

It is clear that the quantum corrections make significant
qualitative changes to the overall structure of the spacetime, as
can be seen by comparing the $\phi$, $\rho$, and curvature scalar
plots for classical and quantum
 cases. But one
would expect such differences from looking for example at the
expressions for the
curvature scalar as a function of $\omega$, $\phi$, and $\rho$.

The curvature scalar is given by
\begin {equation}
R=-8e^{-2\rho}D\rho.
\end{equation}
The field equations can be used to rewrite this expression as
\begin {equation}
R_{cl}=4(2\omega-1)-16\omega s{\phi'}^2 e^{-2\rho},
\end{equation}
\begin {equation}
R_{q}=\frac {4(2\omega-1)-16\omega s{\phi'}^2 e^{-2\rho}}
{P(\epsilon,\phi)^2- \omega e^{2\phi}}
\label{eq:rq}
\end{equation}
in the classical and quantum-corrected cases respectively.
Therefore at weak coupling $(\phi<<0)$, the two quantities may be
 approximately equal,
but clearly, there are large effects near the critical value.
Indeed, several
examples will be seen of the `semi-classical' type of
 singularity which
happens when $\phi$ hits
the critical value.
Let us define the {\it semi-classical singularity} to be one
 where the
dilaton field is finite. The coordinates may or may not diverge at
this point, but this clearly depends on the coordinate system.
The key difference between  this singularity and the classical ones
is simply that the dilaton field no longer diverges there.
 Note that if $\omega\leq 0$ there can be
no semi-classical singularities, although in some cases there are still
 qualitative
differences in causal structure due to the corrections.

\vbox{
\[
\begin{array}{c}
\epsfxsize=0.6\displaywidth
\epsfbox{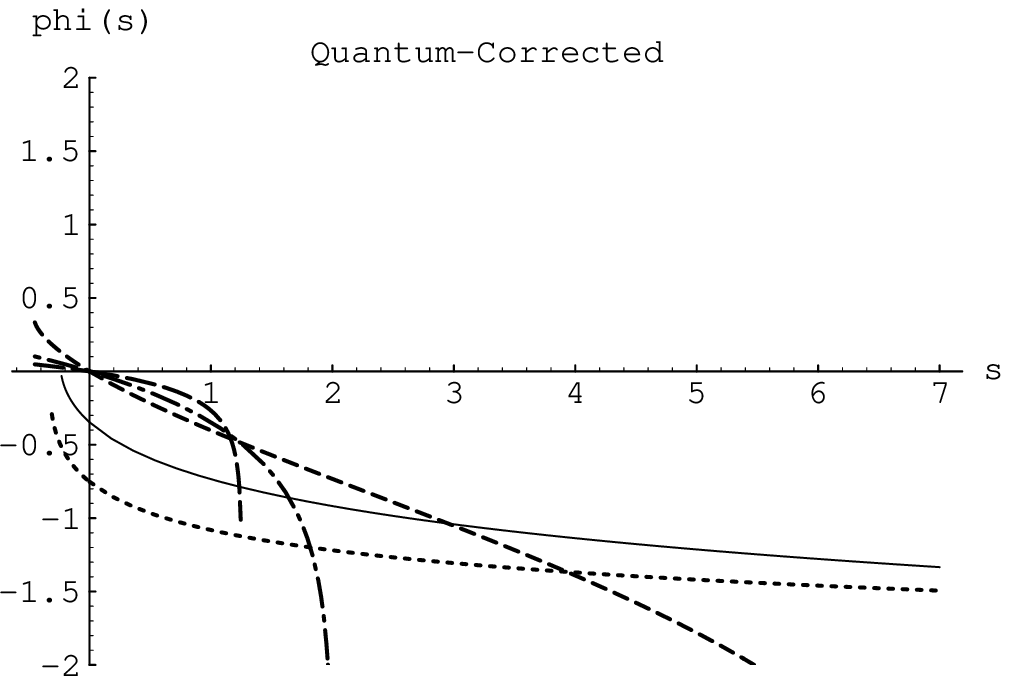}\\
\end{array}\]
FIG. 4. The dilaton at the semi-classical level for
subcritical initial value.}

\subsection{Quantum Solutions}
In the work of \cite{BGHS}, given in their
FIG. 2, $\phi_{cr}=-2$, and
$\kappa=e^4$.  One should note again that this singularity
occurs at finite $\phi$ and $\rho$. Indeed for this choice of
$\kappa$, the singularity is actually at
weak coupling $g_s=e^{-2}$.
The same equations have been solved here in the case
$\omega=1$, with the same initial conditions but different $\kappa$
and initial value of dilaton. The results are qualitatively
identical as expected.

In the new plots given here, $\kappa$ and $\lambda$ have been
scaled out of the  equations by
using field redefinitions of the variables $\rho$ and $\phi$.
 For the quantum case, in general, three sets of graphs for each
value of
$\omega$ are needed, whereas in all the classical plots, $\phi$
ranges  $(-\infty,+\infty)$. In FIG. 4, for $\omega=1$, however,
 $\phi<0$.  Witten\cite{W},
regarded the dilaton field as a
coordinate-independent measure of an observers position. Viewing
it as such, one might believe some of space to have  been omitted as
the region
$\phi\ge 0$ does not exist.
 This is why we may need to plot more than one graph
for each  $\omega$. Then, combining
the solutions, $\phi$ ranges $(-\infty,+\infty)$ as in the classical
case. Although there
will be a singularity as $\phi$ goes through the critical value,
the metric will
be finite there, unlike at the classical counterparts. It is not
sensible, in the
context of semi-classical dilaton gravity, to talk about how objects
 could pass through
the singularity. However,  one can find solutions for values of
$\phi$ above
critical, and thus approach the singularity from
`either side' as far as the dilaton is concerned.

As one approaches the critical value however, the equations which
are derived here from the action(\ref{eq:gen}) no longer represent
the quantum theory of the action. This is because the
graviton-dilaton loops become comparable to the large N matter field
corrections. Thus, it is not clear how one should interpret the
semi-classical singularity. One cannot make definite statements because we
do not have a perturbative expansion or exact theory which indicates
whether or not it persists. On either side of the singularity,
however, the equations should be reliable.

In \cite{BGHS}, purely super-critical solutions were
considered interesting, and such a solution, that of constant
curvature space, was presented.
There exist four-dimensional extremal black hole solutions for
which the asymptotic dilaton value is super-critical\cite{BAN}.
For this reason, this author believes that it is useful to include
the super-critical solutions here, even if some of the section on
planar general relativity is superceded, because
the quantum theory may turn out not to have a
well-defined evolution through the critical value.

 The semi-classical appearance of a
singularity is a limitation. It cannot be
integrated through
numerically and there is no contact between the sub- and
super-critical dilaton regions of spacetime.
A smoothed singularity
 should be passed
through by test particles. In that case it seems plausible that
 one consider how to
paste together semi-classical spacetimes which display
singularities which do not
appear at infinitely strong dilaton
coupling. Naturally, an objection is that
one has to reapply boundary conditions for the super-critical region,
and so the solutions may have nothing to do with the subcritical
ones. However, it is plausible to apply equivalent conditions. This
is supported by the fact that the dilaton field is continuous
across the critical line when one considers attaching
sub- and super-critical solutions.

As long as one has a classical
spacetime picture,
it would seem difficult to go further than this. However, it is
precisely this type of operation that one has in mind for
evaporating black holes which develop baby universes. The
spacelike boundary is removed and replaced with a region to
the future, disconnected from the external space.

 The quantum-corrected cases are commented upon individually in
the following.

\vbox{
\[
\begin{array}{c}
\epsfxsize=0.6\displaywidth
\epsfbox{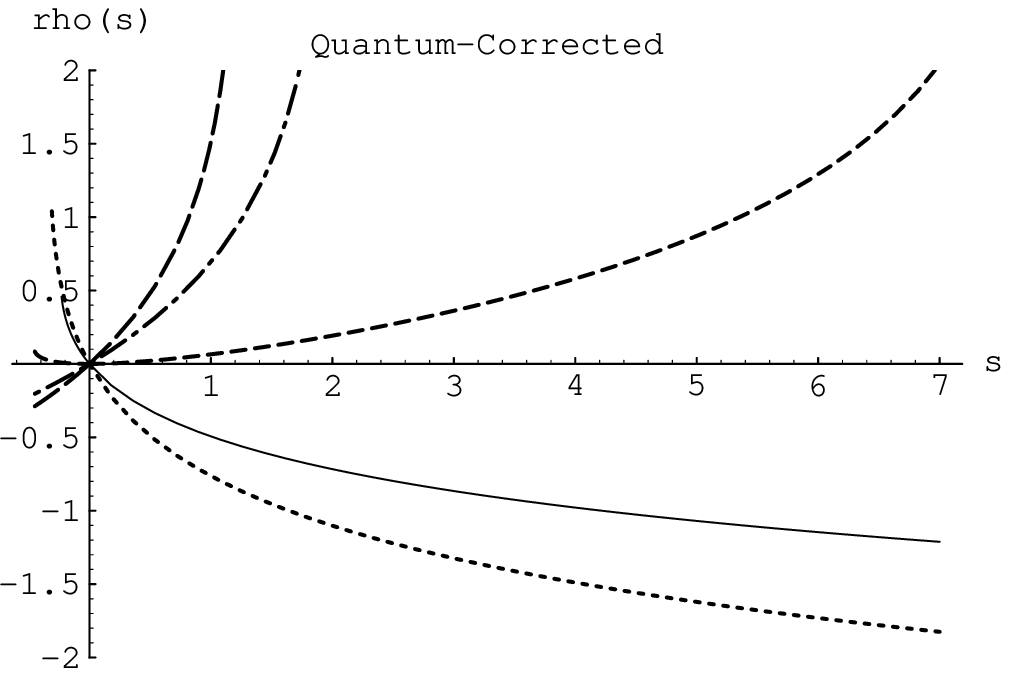}\\
\end{array}\]
FIG. 5. The conformal factor at the semi-classical level for
subcritical initial value.}

\subsection{Sub-Critical Solutions}

Since the critical value is at $\phi_{cr}=-\frac 1 2\log \omega$, the
lowest two cases, $\omega=0,-1$ do not have such
a critical value. But since there is a change of initial gradient of
$\phi$ when $\phi_{0}=\frac 1 2 \log 2$, we use this to divide the
cases, whilst the other cases, $\omega=\frac 1 2, 1, \frac 3 2$ are
genuinely subcritical in the FIG. 4-6.
Generally speaking, the subcritical corrected cases resemble their
classical partners at weak coupling, which is why sub-critical
solutions have received more attention.
The differences become clearer
as the critical value is approached from below.

\vspace{0.3in}
\centerline{\bf A. \ $\omega=-1; \phi_0=0$ }
\smallskip
This case covers the qualitative behaviour for all
$0>\omega>-\infty$.
At strong coupling toward negative $s$, the curvature goes to minus
infinity at $s=-\infty$. To the right, $\phi\to -\infty$ and
$\rho\to\infty$ at finite s,  whilst the curvature
is always negative and finite. Thus we have a timelike
{\it classical type} singularity
to the left, and timelike infinity to the right, rather like  the
extremal Reissner-Nordstrom spacetime in four dimensions. This
does not  differ globally from the classical case.

\vspace{0.3in}
\centerline{\bf B. \ $\omega=0$; $\phi_0=0$}
\smallskip
The classical and quantum types of
singularity coincide, because the place where the quantum
singularity occurs in the
case of $\omega=0$ is at infinity, as in the classical case.

 This is again anti-de Sitter space. The global structure is the same
as for the classical case, though the dilaton goes to infinity
toward the left in a logarithmic fashion in the corrected case.
Lemos and S\` a  showed that the classical Jackiw-Teitelboim theory
contains a non-singular black hole, and this would appear to apply
also to the quantum-corrected case. This black hole has zero
radiation density as $\omega=0$, and zero temperature,  as Cadoni and
Mignemi showed\cite{CM}.

\vbox{
\[
\begin{array}{c}
\epsfxsize=0.6\displaywidth
\epsfbox{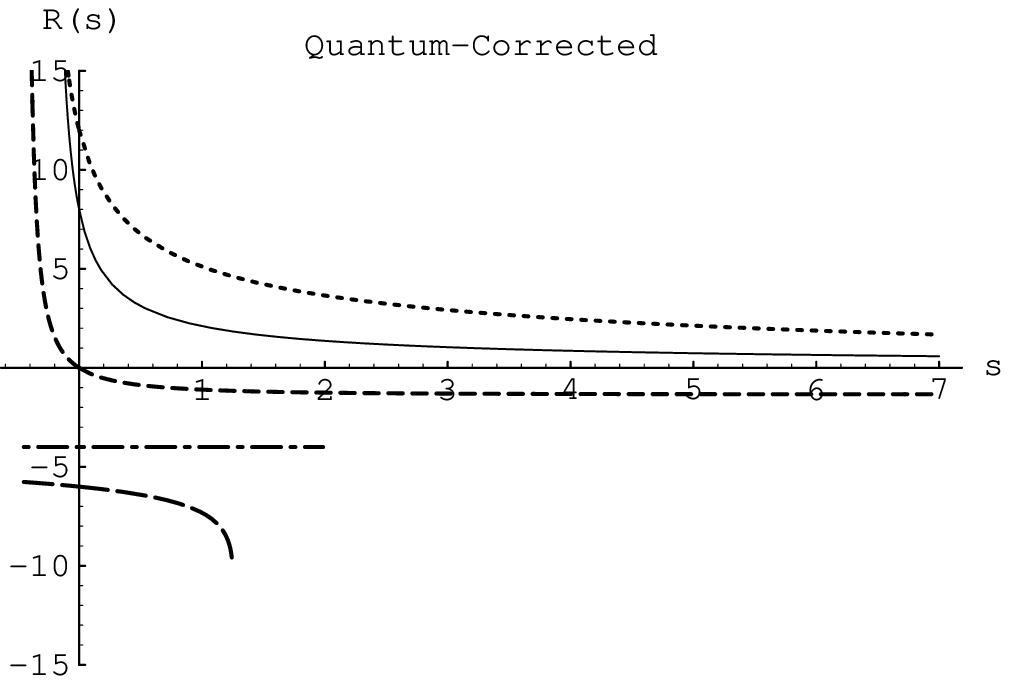}\\
\end{array}\]
FIG. 6. The curvature at the semi-classical level for
subcritical initial value.}
\vspace{0.3in}
\centerline{\bf C. \ $\omega=\frac 1 2, 1, \frac 3 2 $}
\smallskip
For $\omega=\frac 1 2$, the critical value coincides with the
important value $\phi=\frac 1 2\log 2$, (see \ref{eq:derphi}).

Each of these cases has a spacelike singular black hole to the
left, as do higher values of $\omega$. The singularities occur at
finite $\phi$, so the $\phi$(FIG. 6) and $\rho$(FIG. 7) graphs are
truncated behind the horizon. In the asymptotic region to the right,
the behaviour is as classically as expected at weak coupling for
sub-critical plots, quantum corrections being small; the cases
$\omega=\frac 1 2$ and $\omega=1$ are asymptotically anti-de Sitter
and flat respectively, whereas for $\omega>1$ the curvature goes to
minus infinity logarithmically, giving a timelike singularity,
 though
this fact is not clear from FIG. 6, but can easily be confirmed
by plotting to larger $s$.
This zero coupling timelike singularity seems somewhat
unphysical.  The $\omega=1$ case agrees with the work of
Birnir {\it et al}\cite{BGHS}.
{}From the generalised asymptotic expansion given in\cite{SHE},
\begin{equation}
\rho=\log{2b} -
{K+L\log s \over s^{2b}} +
\Big[{K+L\log s \over s^{2b}}\Big]^2 + ...  \label{eq:rho}
\end{equation}
where $K$,$L$ and $b$ are arbitrary constants to be determined,
and
\begin{equation}
\phi = \rho - b\log s - c.  \label{eq:phi}
\end{equation}
Using(\ref{eq:mass}), in terms of the coordinate $2x=b\log s+2c$,
where c is a constant,
 the ADM mass is
\begin{equation}
 M = e^{2c}(K + {1\over{2b}}L(x+\alpha))_{x\to \infty}.
 \label{eq:infty}
\end{equation}
where $\alpha$ is a constant.
The mass is therefore formally infinite, as was seen in\cite{BGHS}.

It is because of the correspondence with classical theory that we
regard the sub-critical case as  physically interesting.
 But this has
meant that the super-critical case has been left uninvestigated,
and
the semi-classical singularity mysterious. In the following
`the other side'
of the singularity, on which the dilaton is super-critical, and
which was originally termed the `Liouville Region',
is considered.

\vbox{
\[
\begin{array}{c}
\epsfxsize=0.6\displaywidth
\epsfbox{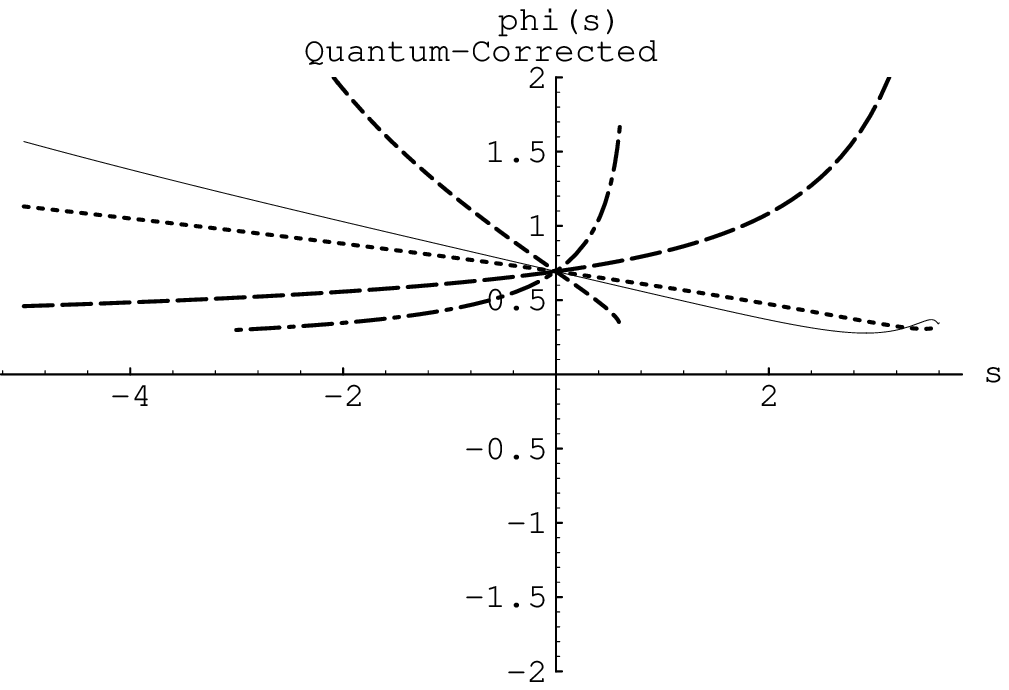}\\
\end{array}\]
FIG. 7. The dilaton at the semi-classical level for
super-critical initial value.}

\subsection{Super-Critical Solutions}

The super-critical initial value is taken to be $\phi_{0}=\log 2$
in the following five cases, which are given in FIG.s 7-9.
 For the
cases $\omega=1,\frac 3 2$, there is another region between the
critical value and $\phi_{0}=\frac 1 2\log2$ in which the dilaton
remains confined. This additional super-critical pair of solutions
is given in FIG.s 10-12 for completeness.

\vbox{\[
\begin{array}{c}
\epsfxsize=0.6\displaywidth
\epsfbox{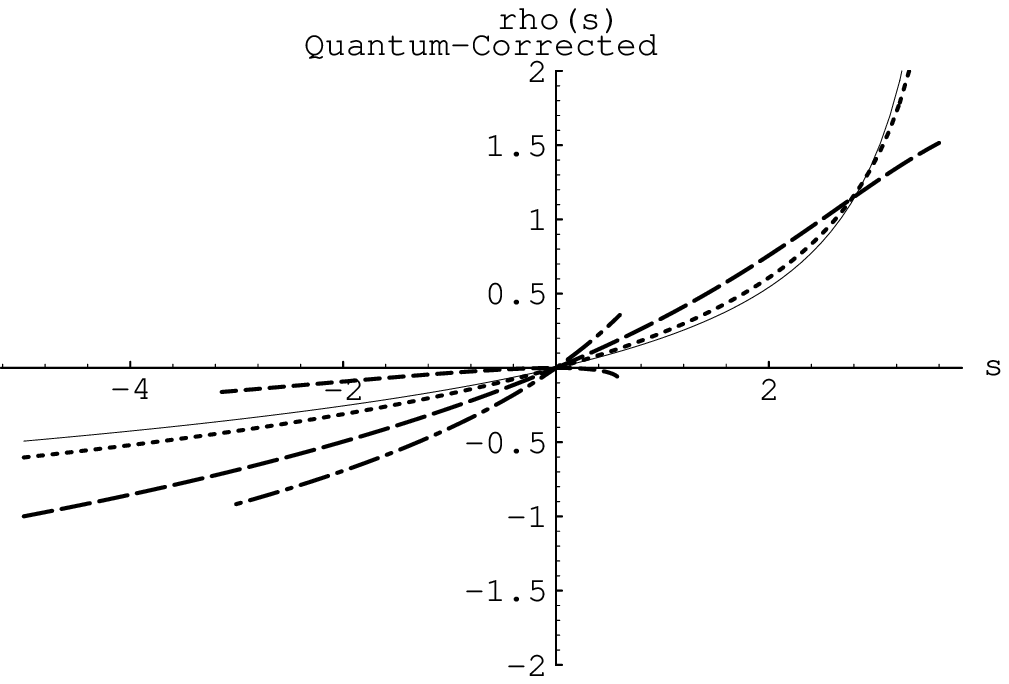}\\
\end{array}\]
FIG. 8. The conformal factor at the semi-classical level, for
supercritical initial value.}

\vspace{0.3in}
\centerline{\bf A. \ $\omega=-1,0$}
\smallskip
Now the strong coupling region is toward positive $s$, and $\phi$
decreases slowly at negative $s$. The space is again anti-de Sitter
for $\omega=0$, while for $\omega=-1$ the curvature varies from
zero at positive $s$,  to $-4$ at negative $s$, where the spacetime
approaches the anti-de Sitter space of the $\omega=0$ case.

\vspace{0.3in}
\centerline{\bf B. \ $\omega=\frac 1 2$}
\smallskip
This is the only case which has a spacelike singularity; the other
cases are disconnected from their critical values in this region. In
particular, the critical value is  $\phi_{cr}=\frac 1 2 \log 2$.
To the left $\phi$ diverges but the curvature is negative and
slowly varying, while to the right, there is a singularity hidden by
a horizon. We return to this case below.

\vbox{\[
\begin{array}{c}
\epsfxsize=0.6\displaywidth
\epsfbox{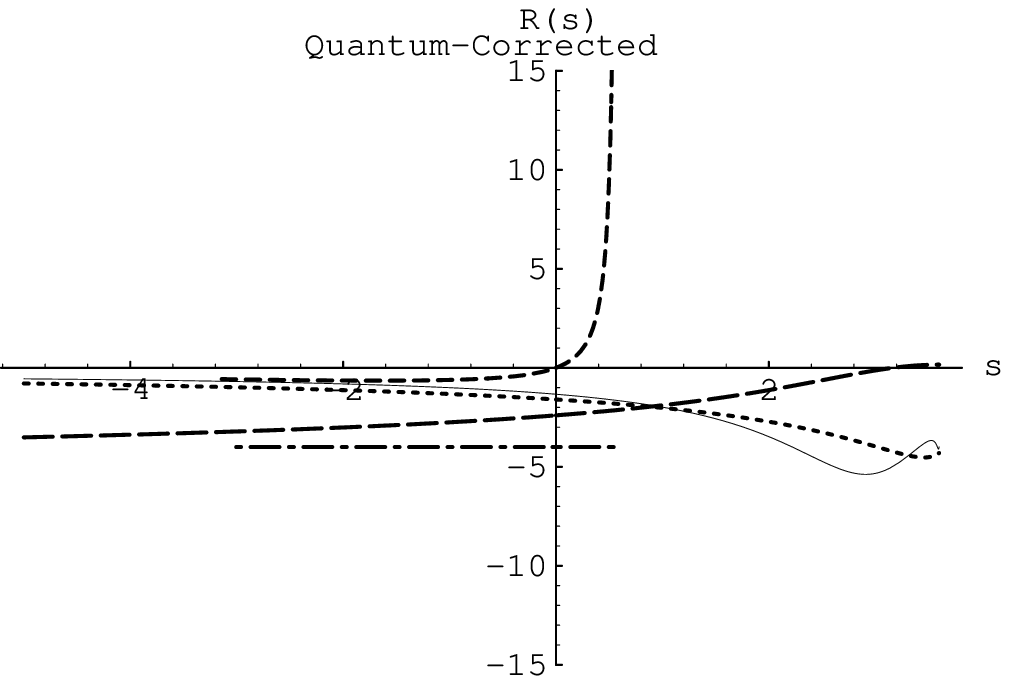}\\
\end{array}\]
FIG. 9. The curvature at the semi-classical level, for
supercritical initial value.}

\vspace{0.3in}
\centerline{\bf C. \  $\omega=1,\frac 3 2$;
$\phi_0>\frac 1 2 \log 2$ }
\smallskip
These have finite curvature everywhere in this region.
Toward negative
$s$, at strong coupling, the space is asymptotically flat, while
toward positive $s$, as $\phi\to\frac 1 2\log 2$, the curvature
$R\to -4$. These results are in FIG.s 7-9.

\vspace{0.3in}
\centerline{\bf D. \ $\omega=1,\frac 3 2$;
$\phi_{cr}<\phi_0<\frac 1 2 \log 2$ }
\smallskip

One can also consider the region immediately above the singularity
as far as $\phi$ is concerned for these parameter values. The two
examples are qualitatively similar. They have a timelike
singularity at negative $s$ as the dilaton descends toward the
critical value, and have approximately constant negative
curvature at positive $s$, where $\phi\to\frac 1 2\log 2$ and
$\rho$ diverges. These comments are given graphically in FIG.s
10-12.

\vbox to \vsize{\[
\begin{array}{c}
\epsfxsize=0.6\displaywidth
\epsfbox{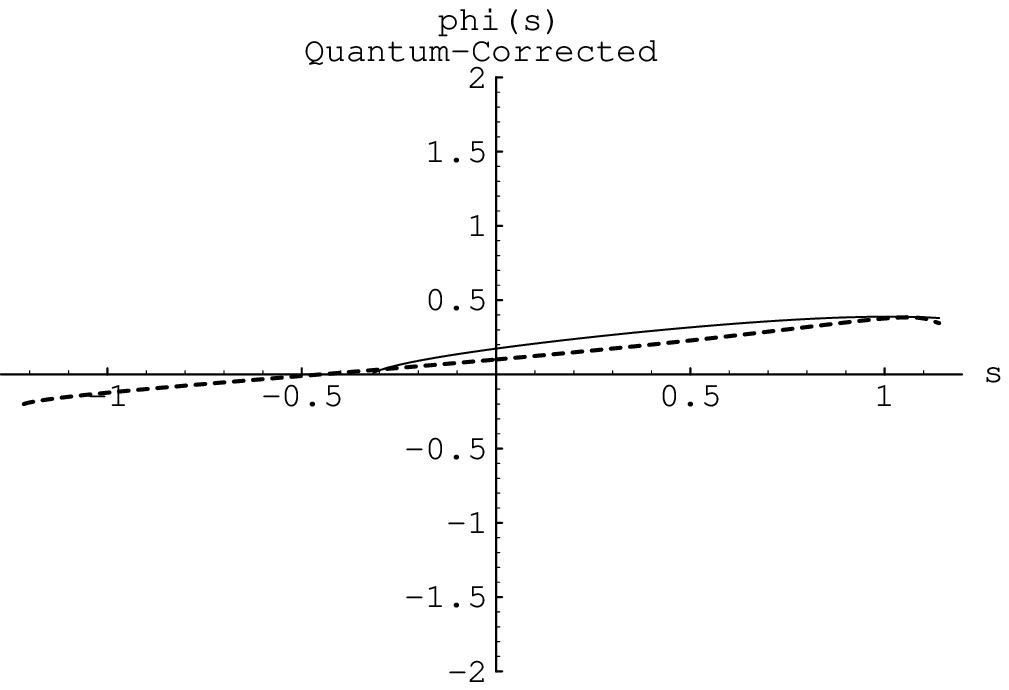}\\
\end{array}\]
FIG. 10. The dilaton for
$\omega=1,\frac 3 2$ at the semi-classical level, for
initial value which is above critical but below $\frac 1 2 \log 2$.
\bigskip
\[
\begin{array}{c}
\epsfxsize=0.6\displaywidth
\epsfbox{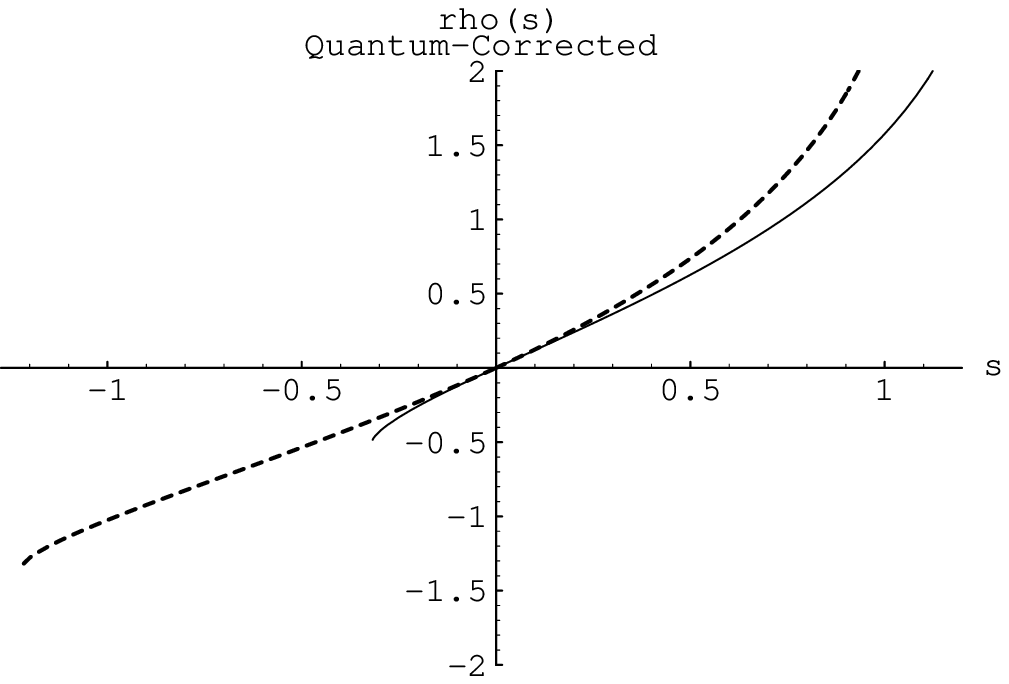}\\
\end{array}\]
FIG. 11. The conformal factor for
$\omega=1,\frac 3 2$ at the semi-classical level, for
initial value which is above critical but below $\frac 1 2 \log 2$.\vfil}

\vbox{\[
\begin{array}{c}
\epsfxsize=0.6\displaywidth
\epsfbox{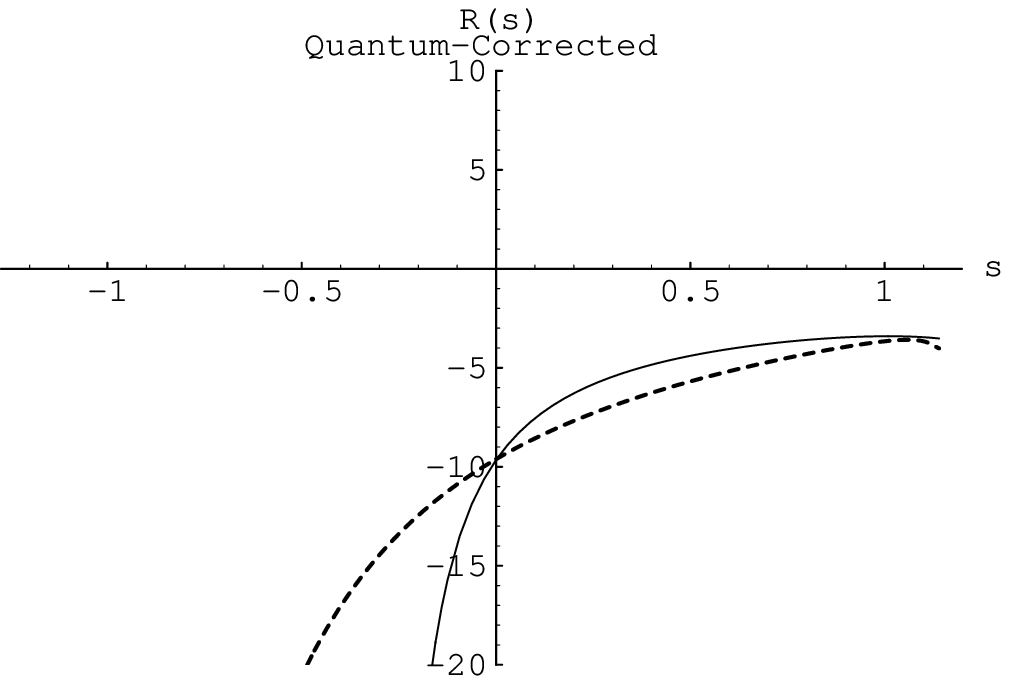}\\
\end{array}\]
FIG. 12. The curvature for
$\omega=1,\frac 3 2$ at the semi-classical level, for
initial value which is above critical, but below $\frac 1 2 \log 2$.}

\subsection{Planar General Relativity}

The classical core of this case has been discussed in\cite{LEM}.
The classical solution given earlier is actually a solution
of general relativity.
In this case, there are only two regions necessary to
cover all values of $\phi$ along the real line, because the critical
value coincides with the value $\phi=\frac 1 2\log 2$, which was
discussed earlier. Generally, in such cases where there exists
a finite region $\phi_{cr}<\phi<\frac 1 2 \log 2$,
there is a solution in which the dilaton is
confined between these outside values so that
a third initial value will be necessary for the dilaton to range
the full real line.
Consider the results given in FIGS. 4 and 7.
 The curve $\omega=\frac 1 2$ in FIG. 7 increases monotonically
from
the asymptotic region at the right until it reaches $\phi_{cr}=
\frac 1 2\log 2$ when there is a singularity and the integration
breaks down. In FIG. 4, the curve descends monotonically to this
 value
 and has the same gradient there. If one were to attach these two
curves, the dilaton would be a continous monotonic function
{\it through} the singularity. There would be a small discontinuity
 in the conformal factor there. For other, asymetric initial
dilaton values, the dilaton curve is no longer smooth, but it
remains monotonic and piecewise continuous.
The reader is reminded, however, that the equations are not valid at
the singularity, so this point may not be important in a fuller
theory.

The Penrose diagrams for the
extended
spacetime corresponds to the two sets of solutions are given in
FIG. 13. Dashed regions are copies of their undashed counterparts.

\setlength{\unitlength}{0.0050in}%
\hspace{60mm}
\begin{picture}(160,520)(320,260)

\put(680,440){\line( 0, 1){160}}
\put(520,440){\line( 0, 1){160}}
\put(680,440){\line(-1, 1){160}}

\put(520,440){\line( 1, 1){160}}

\put(520,600){\line( 1, 0){160}}
\put(523,597){\line( 1, 0){154}}
\put(520,440){\line( 1, 0){160}}
\put(523,443){\line( 1, 0){154}}
\put(650,520){\makebox(0,0)[lb]{\raisebox{0pt}[0pt][0pt]{I}}}
\put(595,565){\makebox(0,0)[lb]
{\raisebox{0pt}[0pt][0pt]{II}}}
\put(595,465){\makebox(0,0)[lb]
{\raisebox{0pt}[0pt][0pt]{II'}}}
\put(530,520){\makebox(0,0)[lb]
{\raisebox{0pt}[0pt][0pt]{ I'}}}
\put(530,400){\makebox(0,0)[lb]
{\raisebox{0pt}[0pt][0pt]{$\phi=\frac 1 2 \log 2$}}}
\put(685,500){\makebox(0,0)[lb]
{\raisebox{0pt}[0pt][0pt]{ $\phi=+\infty$}}}
\put(540,350){\makebox(0,0)[lb]
{\raisebox{0pt}[0pt][0pt]{ Case 2.}}}

\put(380,440){\line( 0, 1){160}}
\put(220,440){\line( 0, 1){160}}
\put(380,440){\line(-1, 1){160}}

\put(220,440){\line( 1, 1){160}}

\put(220,600){\line( 1, 0){160}}
\put(223,597){\line( 1, 0){154}}
\put(220,440){\line( 1, 0){160}}
\put(223,443){\line( 1, 0){154}}
\put(350,520){\makebox(0,0)[lb]
{\raisebox{0pt}[0pt][0pt]{ I}}}
\put(295,565){\makebox(0,0)[lb]
{\raisebox{0pt}[0pt][0pt]{II}}}
\put(295,465){\makebox(0,0)[lb]
{\raisebox{0pt}[0pt][0pt]{II'}}}
\put(230,520){\makebox(0,0)[lb]
{\raisebox{0pt}[0pt][0pt]{ I'}}}
\put(230,620){\makebox(0,0)[lb]
{\raisebox{0pt}[0pt][0pt]{$\phi=\frac 1 2 \log 2$}}}
\put(100,500){\makebox(0,0)[lb]
{\raisebox{0pt}[0pt][0pt]{$\phi=-\infty$}}}
\put(240,350){\makebox(0,0)[lb]
{\raisebox{0pt}[0pt][0pt]{ Case 1.}}}

\end{picture}

\centerline{FIG. 13: Penrose diagrams of
quantum-corrected planar general relativity.}
\vspace{0.2in}

At the spacelike singularities of both boxes,
$\phi=\frac 1 2\log 2$. One could draw a line from the
timelike infinity in case 1., region I.  where $\phi=-\infty$,
through the singularity, where one identifies with a point on the
lower singularity of case 2., and on through region II',
out to timelike infinity in region I'  where $\phi\to\infty$.
A plausible pasting together of the two spacetimes would be to
place the box corresponding to case 2. on top of that of case 1. and
identify the singularities where the dilaton is $\frac 1 2\log 2$.
This is very literally `toy modelling'.
The interpretation of the resulting single diagram is open
to debate. As expected, there is now a singularity at
finite coupling in the {\it middle} of the diagram.
 Semi-classically, the singularity is final.
 However, in quantum gravity, the singularity may be
smoothed, and test particles may be able to pass through the
region of high curvature. The diagram here suggests a strong
curvature wormhole shrouded on either `side' by an horizon, at which
the curvature goes through zero, and is asymptotically
anti-de Sitter.

An observer who begins in the asymptotic region I of case I could
avoid the wormhole by constantly accelerating
immediately to timelike infinity, when $\phi=-\infty$,
 staying in region I. Alternatively, he might
remain stationary, in which case he would pass through the
wormhole at $\phi=\frac 1 2 \log 2$, after which he could
constantly accelerate so that he reached another timelike infinity
in the second box, where $\phi=\infty$.

\section{Conclusions}

A general, two-dimensional model has been considered and solved
numerically for static, equilibrium solutions.
There are many
configurations which depend on both the value of the
parameter $\omega$  and
on the initial value of the dilaton field at the origin.

The classical solutions found, bore out the results of
\cite{LEMSA},
  and the semi-classical $\omega=1,\phi<\phi_{cr}$ case,
those of \cite{BGHS}.
The extreme Reissner-Nordstrom type solution $\omega=-1$, the
Schwarzschild type black hole $\omega=1$, of low-energy string
theory, the black hole with timelike anti-de Sitter infinity
$\omega=\frac 1 2$, of planar general relativity, the
spacetime with both timelike and spacelike singularities,
$\omega=\frac 3 2$, and the non-singular Jackiw-Teitelboim
black hole $\omega=0$, were all seen both classically and at
the semi-classical level, where they in general represented
black holes in the Hartle-Hawking equilibrium state.
These sub-critical solutions are the most clearly physically
interesting solutions since they correspond with their classical
counterparts and four-dimensional analogues.
Of all these
solutions, the unique parameter value which yields a solution
which has everywhere positive curvature and is asymptotically
flat, is that of string theory, $\omega=1$.

Super-critical solutions for all cases were
also found, and for ($\omega=\frac 1 2$), it appeared plausible
to paste the super-critical to the sub-critical solution.
Then, as in classical theories, the dilaton
ranges the full real
line and is continuous across the singularity.
This construction is not
possible in classical
theory since in that case, the singularity always occurs at
divergent dilaton field.

At the semi-classical level,
the two regions are
still
divided by a curvature singularity.
This singularity was not expected originally\cite{CGHS}, and
was met with puzzlement when discovered in subsequent work
\cite{RSTA,BAN,BGHS}.
In the {\sl RST} model, the singularity is
taken seriously as a `central' boundary, analogous to the origin
in Schwarzschild spacetime. However, it
is known that there are energy conservational problems at the
endpoint\cite{PS}, which may be related to this potential
misinterpretation.
The equations which generate the singularity become
inappropriate in its vicinity, but
one can still consider
sub-critical and super-critical solutions independently.

The singularity is a modification to classical theory which
may or may not go away in quantum gravity,
or is generically spurious.
Birnir {\it et al}\cite{BGHS} discussed the possibility of sailing
through this mild singularity.
Horowitz and
Marolf \cite{HM} have recently discussed the behaviour of quantum
test particles which have well-defined motion even in singular
spacetimes.

\newpage

\section{Acknowledgements}
I thank Stephen Hawking and Gary Gibbons
for reading the paper, and for conversations and suggestions.

\end{document}